\documentclass[11pt]{cernrep}
\usepackage{graphicx}
\usepackage{here}

\def\lsim{\mathrel{\mathop
  {\hbox{\lower0.5ex\hbox{$\sim$}\kern-0.8em\lower-0.7ex\hbox{$<$}}}}}
\def\gsim{\mathrel{\mathop
  {\hbox{\lower0.5ex\hbox{$\sim$}\kern-0.8em\lower-0.7ex\hbox{$>$}}}}}

\begin{document}

\newcommand{\ii}{\'{\i}}
\newcommand{\half}{{1\over2}}
\newcommand{\halfpi}{{1\over2\pi}}
\newcommand{\ie}{{\em i.e.}\ }
\newcommand{\eg}{{\em e.g.}\ }
\newcommand{\efolds}{$e$-folds\ }
\newcommand{\Mp}{M_{\rm P}}
\newcommand{\Aloop}{{\cal A}_{\rm 1-loop}}
\newcommand{\Vloop}{V_{\rm 1-loop}}
\newcommand{\trh}{T_{\rm rh}}
\newcommand{\ANS}{{\cal A}_{\rm NS-NS}}
\newcommand{\ARR}{{\cal A}_{\rm R-R}}
\newcommand{\Tr}{\rm Tr}
\newcommand{\Lns}{{\rm L}_0^{\rm (NS)}}
\newcommand{\Lr}{{\rm L}_0^{\rm (R)}}

\newbox\pippobox

\title{INFLATION FROM BRANES AT ANGLES}
\author{Juan Garc\'\i a-Bellido}
\institute{Dept. de F\ii sica Te\'orica C-XI,
Universidad Aut\'onoma de Madrid, Cantoblanco 28049 Madrid, Spain}
\maketitle
\begin{abstract}
In this lecture I will review the present status of inflation within
string-based brane-world scenarios. The idea is to start from a
supersymmetric configuration with two parallel static Dp-branes, and
slightly break the supersymmetry conditions to produce a very flat
potential for the field that parametrises the distance between the
branes, \ie the inflaton field. This breaking can be achieved in
various ways. I will describe one of the simplest examples described
in Ref.~[1]: two almost parallel D4-branes in a flat compactified
space, with a small relative angle between the branes, softly breaking
supersymmetry. If the breaking parameter is sufficiently small, a
large number of \efolds can be produced within the D-brane, for small
changes of the configuration in the compactified directions. Such a
process is local, \ie it does not depend very strongly on the
compactification space nor on the initial conditions. Moreover, the
breaking induces a very small velocity and acceleration, which ensures
very small slow-roll parameters and thus an almost scale invariant
spectrum of metric fluctuations, responsible for the observed
temperature and polarization anisotropies in the microwave
background. Another prediction of the model is the small amplitude of
the gravitational wave spectrum, which could be undetectable in the
near future. Inflation ends as in hybrid inflation, triggered by the
negative curvature of the string tachyon potential.
\end{abstract}

\section{Introduction}

Inflation is a paradigm in search of a unique model, that may eventually
be described within a complete theory of particle
physics~\cite{Lindebook}. For the moment we just have the basics or
fundamentals of inflation, and do not just have a single (definitive)
realization of inflation, but many. So, how do we construct a consistent
model of inflation within particle physics? First, we have to provide an
effective scalar field, the inflaton; but what is the nature of this
inflaton field?  Then we have to give a dynamics for this field, some
effective potential that is sufficiently flat for inflation to proceed;
but how do we attain such flatness? Furthermore, the amplitude and tilt
of (scalar and tensor) metric fluctuations should agree with the
observed CMB temperature and polarization anisotropies, as well as the
LSS matter distribution seen by deep galaxy redshift surveys. These
observations suggest an approximately scale invariant spectrum of
adiabatic Gaussian curvature perturbations with an amplitude of a few
parts in $10^5$.

Over the years there have been a handful of particle physics
candidates of inflation~\cite{LythRiotto} based on a variety of
concepts: a) {\em symmetry breaking} (to trigger the end of
inflation), in models of hybrid inflation~\cite{hybrid}; b) {\em
supersymmetry} (in order to cancel dangerous polynomial loop
corrections) which gives rise to reasonably flat directions like in D-
or F-term inflation~\cite{LythRiotto}; and c) {\em string theory} (in
search of a consistent theory of gravity at the quantum
level~\cite{p98}), which generically gives slow-roll parameters of
order one, and thus makes sufficient inflation difficult to realize.

Recently, the proposals of
Refs.~\cite{dt98,dss01,bmnqrz01,hhk01,dhhk02,klps01,ks01,st01} have
managed to derive some of the inflationary properties from concrete
non-supersymmetric stringy (brane) configurations. A new development
started in the mid-nineties, within $d=10$ superstring theory, with
the ``discovery'' of the D(p)-branes, i.e. non-perturbative extended
(p+1)-dimensional objects on which fundamental strings could live
and/or attach their ends to. In the presence of branes, strings can
satisfy two distinct boundary conditions~\cite{p98}:
\begin{equation}
\begin{array}{lr}
{\rm Neumann:} \ & \left.X^\mu(\tau, \sigma)\right|_{\rm brane} = 0\,,\\[2mm]
{\rm Dirichlet:} \ & \left.\partial_\sigma 
X^\mu(\tau, \sigma)\right|_{\rm brane} = 0\,.\\
\end{array}
\end{equation}

\begin{figure}[htb]
\vspace*{-.5cm}
\begin{center}
\includegraphics[width=10cm]{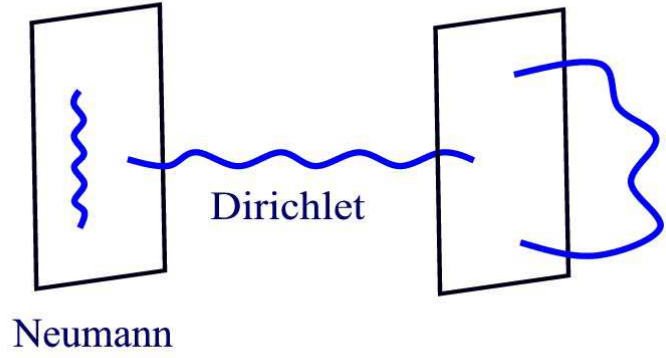}
\vspace*{-1cm}
\caption{This figure represents two Dp-branes with strings
living along the brane, satisfying Neumann boundary conditions, as
well as open strings with both ends attached to one or two branes,
satisfying Dirichlet boundary conditions.}
\end{center}
\label{Dbranes}
\end{figure}

These D-branes are dynamic objects, whose forces arise from the
exchange of open and closed string modes between them. One can compute
exactly (within string theory) the potential corresponding to the
interaction between Dp-Dp or Dp-$\overline{\rm Dp}$
branes~\cite{p98}. There are two possible types of modes
exchanged between branes: Ramond-Ramond (R-R) modes (fermions,
$C_p$-potentials) and Neveu-Schwarz-Neveu-Schwarz (NS-NS) modes
(graviton, dilaton). When supersymmetry is preserved, the NS-NS and
R-R charges cancel and there is no net force between the branes.

In that case, the 1-loop amplitude due to exchange of string modes
cancels exactly, $\Aloop = \ANS - \ARR  = 0$, where~\cite{p98}
\begin{equation}
\Aloop = 2V_{p+1}\int_0^\infty{dt\over2t}\int{d^{p+1}k\over
(2\pi)^{p+1}}\left[\Tr{1+(-1)^F\over2}\,e^{-2\pi t\,\Lns}-
\Tr{1+(-1)^F\over2}\,e^{-2\pi t\,\Lr}\right]\,.
\end{equation}
Thus, two parallel Dp-branes constitute a stable BPS state. The
tree-level potential is solely determined by the sum of the branes'
tensions, $V_0 = 2T_p =$ constant, and therefore generates no force.

In this context, it was proposed by Dvali and Tye~\cite{dt98} the
idea of {\em brane inflation}, in which the role of the inflaton field
is played by the interbrane separation $y$. They treated the potential
phenomenologically, without actually deriving it from string theory. 
Just assuming Gauss law for the massless modes propagating between
branes, they wrote the potential at large distances as
\begin{equation}
V(y) \approx V_0 - {B\over y^{d_\bot - 2}}\,,
\end{equation}
where $d_\bot = 10 - (p+1)$ is the number of compact dimensions
transverse to the branes and $B$ is a certain coefficient, in
principle derivable from string theory. With this potential, inflation
occurs, but requires unnatural fine tuning of $B\ll 1$ in order to have
sufficient number of \efolds.

One of the main interests of dealing with these non-perturbative
objects in string theory is that they are extended objects of
dimension (p+1) and, in principle, our (3+1) world could be embedded
in them. If the forces between these objects give rise to effective
potentials for certain scalars, one could then produce inflation
within the branes, in the context of string theory, and thus explain
our present homogeneity and flatness. This is what makes the scenario
very attractive.

\begin{figure}[htb]
\vspace*{-.5cm}
\begin{center}
\includegraphics[width=10cm]{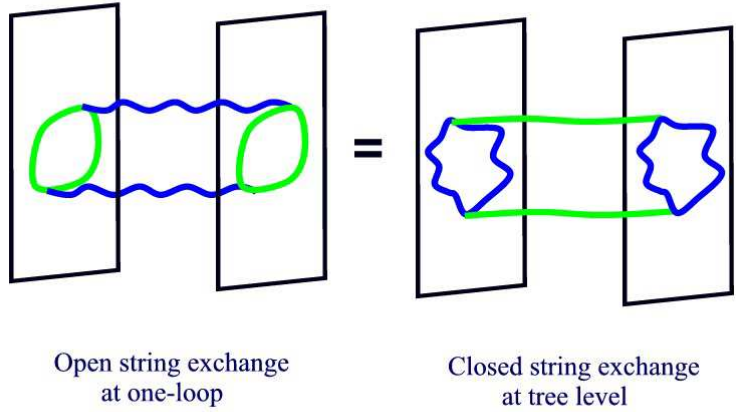}
\vspace*{-0.5cm}
\caption{The interaction between two Dp-branes arise from the
exchange of closed string modes at tree level, which is equivalent
to the 1-loop exchange of open string modes.}
\end{center}
\label{OpenClose}
\end{figure}

\section{Brane-antibrane inflation}

Soon after the proposal of Dvali and Tye, two groups obtained
independently an exact expression for $V(y)$ in the case of maximal
supersymmetry breaking of the BPS, i.e. with a brane and an
antibrane~\cite{bmnqrz01,dss01}. Since the antibrane has opposite RR
charges than the brane, the 1-loop amplitude does not vanish, $\Aloop
= \ANS - \ARR = 2\ANS \neq 0$, and thus there appears a force at
1-loop between the brane and the antibrane. Let us suppose for
concreteness that the brane has dimension (3+1) like our own world,
and thus consider a system of D3-$\overline{\rm D3}$ branes. Such
branes have (3+1) dimensions which are Neumann with respect to the
propagation of the strings embedded in them, and $d_\bot=6$ transverse
compactified dimensions, which are assumed to be fixed
somehow,~\footnote{This is one of the most difficult problems to be
addressed, not only in this scenario but in string theory in general.}
along which the bulk strings satisfy Dirichlet boundary conditions.
An schematic view of this brane-antibrane system can be seen in
Fig.~3. The branes are separated a certain distance 
apart in the compactified space, as represented by the crosses in the
planes $(X_4,\,X_5),\ (X_6,\,X_7),\ (X_8,\,X_9)$.

\begin{figure}[htb]
\begin{center}
\includegraphics[width=12cm]{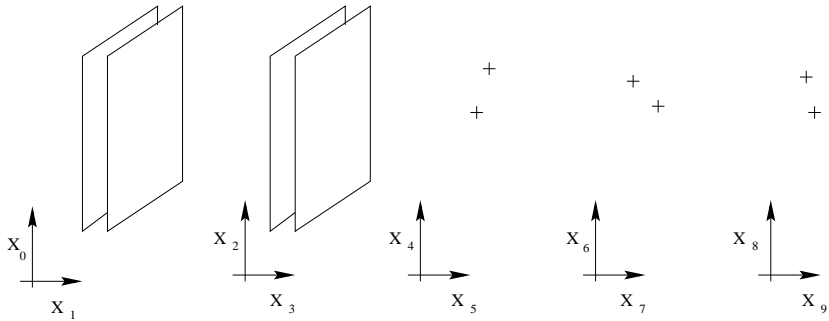}
\caption{This figure represents two D3-branes. The usual 4-dimensional
spacetime is along directions $\{x^0,...,x^3\}$. The branes are
located at particular points on the remaining six compact directions
$\{x^4,...,x^9\}$.}
\end{center}
\label{angleD3}
\end{figure}

At large distances, the 1-loop amplitude is dominated by the term
\begin{equation}
\ANS\propto G_{9-p}(y) \propto {1\over y^{d_\bot-2}} = {1\over y^4}\,,
\end{equation}
where $G_{9-p}(y)$ is the massless propagator in $(9-p)$ dimensions.  As
we will see, the proportionality constant depends on the string theory
couplings and can be calculated explicitly.

On the other hand, at short distances, it is well
known~\cite{p98} that the low energy effective theory of the
brane-antibrane system contains a tachyon whose mass squared becomes
negative,
\begin{equation}
m_T^2 = {y^2\over4\pi^2{\alpha'}^2} - {1\over2\alpha'}\,,
\end{equation}
below a certain distance $y\leq y_c=\sqrt{2\pi^2\alpha'} \equiv
\sqrt{2\pi^2}\ l_s$, of order the string scale. A tachyon signals the
instability of the system and therefore the end of inflation. In fact,
the behavior of the system is very similar to that of hybrid
inflation~\cite{hybrid}.  It is probably the first example of the
generic model of hybrid inflation in the context of string theory.

This simple model, however, has a fundamental flaw: after inflation,
the brane-antibrane disappear completely (they mutually annihilate),
radiating into the bulk, not along the brane, and thus would {\em not}
produce the required reheating of the universe after inflation
responsible for the subsequent stages of Big Bang cosmology.  There
are a few ways out, by including different types of branes of higher
dimension which decay into lower-dimensional branes, see
\cite{bmnqrz01}, or by adding extra degrees of freedom like electric
or magnetic fluxes on the branes, which will prevent their complete
annihilation.

In any case, this model does not produce enough inflation to solve the
flatness and homogeneity problems. The reason is quite simple:
superstring theory give coefficients and couplings that are of order one
in string units, and therefore, the brane and antibrane attract
eachother too strongly and collide before the brane-world has time to
expand sufficiently to solve those problems. Let us see this explicitly
for the case of D3-$\overline{\rm D3}$ branes~\cite{dss01,bmnqrz01}.

The 4-dimensional effective action for the interbrane separation $y$
is given by
\begin{equation}
S = \int d^4x\,\sqrt{-g}\left[\half M_P^2\,R - \half T_3\,
(\partial y)^2 - V(y)\right]\,,
\end{equation}
where $T_3=M_s^4g_s^{-1}/(2\pi)^3$ is the D3-brane tension, and the
4-dimensional Planck mass is given by $M_P^2 = M_s^2(M_s^6V_6)g_s^{-2}$,
with $V_6 \equiv (2\pi R)^6$ the volume of the 6-dimensional
compactified space (here assumed to be a 6-torus of common radius $R$).
The effective potential for the inflaton can be obtained from
string theory (for $p=3$), assuming that the distance between branes is
large compared with the string scale but small compared with the
compactified space ($l_s \ll y \ll 2\pi R$), as
\begin{equation}
V(y) = 2T_p + 4\pi(4\pi^2\alpha')^{3-p}\,G_{9-p}(y) \equiv
M^4\left(1-{\alpha\over\phi^4}\right)\,,
\end{equation}
where $\phi \equiv T_3^{1/2}y$ \ is the canonically normalized inflaton 
field, and \ $\alpha = M^4/32\pi^2$.

It is easy to compute the number of \efolds from this potential,
\begin{equation}
N = {1\over M_P^2}\int{V\,d\phi\over V'(\phi)} = 
{2\pi^2\over3(2\pi)^6}\,\left({y\over2\pi R}\right)^6\,.
\end{equation}
In the approximation in which this potential was computed, $y \ll 2\pi
R$, the number of \efolds is much smaller than one and thus insufficient
for solving the homogeneity and flatness problems. Put another way, in
order to have enough number of \efolds, $N\simeq 60$, one would require
$y > 2\pi R$, and therefore, one would have to take into account the
infinite series of images of the branes (and antibranes) in the
compactified space. This was considered in Ref.~\cite{bmnqrz01} and found to
give rise to an effective potential which could marginally provide
enough number of \efolds, as long as the initial conditions of inflation
were chosen very close to a symmetric configuration. We will not pursue
this issue here, but will now describe an alternative and quite robust
solution for brane inflation that we came up with in Ref.~\cite{brz01}.

\section{Inflation from branes at angles}

In order to overcome the difficulties of constructing successful
inflationary models from brane-antibrane interactions, where
supersymmetry is maximally broken, we considered in Ref.~\cite{brz01}
the possibility of generating the inflaton potential from the
spontaneous breaking of supersymmetry, so that one can smoothly turn
on an interaction between the branes. There are many ways in which
this general idea can be implemented: by a slight rotation of the
branes intersecting at small angles (equivalent, in the T-dual
picture, to adding small magnetic fluxes through the branes), by
considering small relative velocities between the branes, by
introducing orientifolds, etc. In Ref.~\cite{brz01} we considered the
interactions that arise between intersecting branes at a small angle.
If the angle is small then supersymmetry is only slightly broken, so
that the force between the branes is relatively small. It is this flat
potential which will drive inflation, as the branes are attracted to
each other, until a tachyonic instability develops when the branes are
at short distances compared with the string scale. This is a signal
that a more stable configuration with the same R-R total charge is
available, which triggers the end of inflation.

In the following we elaborate on one of the simplest examples: a pair
of D4-branes intersecting at a small angle in some compact
directions~\cite{brz01}. The interaction can be arbitrarily weak by
choosing the appropriate angle to be sufficiently small. The
brane-antibrane system is the extreme supersymmetry breaking case,
where the angle is maximised such that the orientation for one brane
is opposite to the other. The interaction is so strong in this case,
that inflation seems hard to realise~\cite{dss01}. One should take
very particular initial conditions on the system for inflation to
proceed~\cite{bmnqrz01}.  On the other hand, if the supersymmetry
breaking parameter is small, a huge number of \efolds are available
within a small change of the internal configuration of the system, due
to the almost flatness of the potential; in this way the initial
conditions do not play an important role.  Thus, we find that
inflation is robust in systems that are not so far from
supersymmetry preserving ones.

The following table gives us some perspective on the model building
possibilities for inflation in the context of stringy brane
constructions.

\begin{table}[h]\hspace{.5cm}
\begin{tabular}{llll}
$\theta = 0$ & Brane-Brane (parallel) & 
SUSY preserved & No force (static) BPS \cr
$\theta = \pi$ & Brane-Antibrane (antiparallel) &
maximal SUSY breaking & Large attractive force \cr
$\theta \ll \pi$ & Brane-Brane (small angle) &
spontaneous SUSY breaking & Weak attractive force
\end{tabular}
\end{table}

The aim is to obtain an almost flat potential for the moduli field
describing the interbrane separation. One of the interesting features
of branes in string theory is that they can wrap around certain
cycles in the compactified space. The wrapping has associated with it
an angle $\theta$ and a length $L$, see Fig.~5. These
parameters will characterize the inflaton potential.

\section{Description of the model}

In our paper~\cite{brz01}, we considerd type IIA string theory on
${\cal R}^{3,1}\times T^6$, with $T^6$ a (squared) six torus.  We put
two D4-branes expanding 3+1 world-volume dimensions in ${\cal
R}^{3,1}$, with their fourth spatial dimension wrapping some given
1-cycles of length $L$ on one of the $T^2$ in $T^6$. In
Fig.~4 we have drawn a concrete configuration.

\begin{figure}[htb]
\begin{center}
\includegraphics[width=12cm]{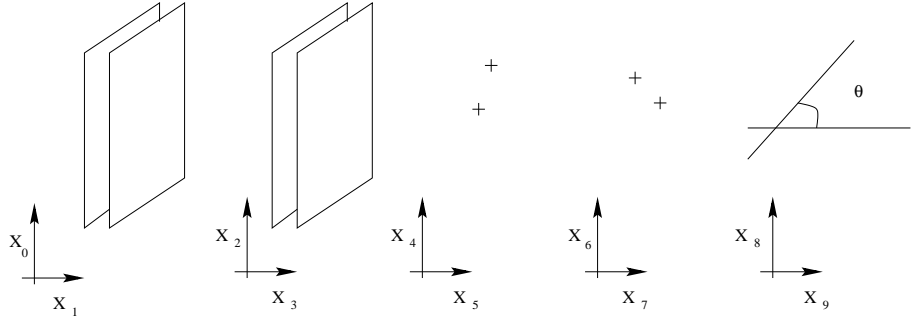}
\caption{This figure represents two D4-branes at an angle $\theta$. The
usual 4-dimensional spacetime is along directions $\{x^0,...,x^3\}$. The
branes are located at particular points on the four compact directions
$\{x^4,...,x^7\}$. Finally, they wrap different cycles on the last two
compact directions, $x^8$ and $x^9$, and intersect at a given angle
$\theta$.}
\end{center}
\label{angleD4}
\end{figure}

If both branes are wrapped on the same cycle and with the same
orientation, we have a completely parallel configuration that preserves
sixteen supercharges, \ie ${\cal N}=4$ from the 4-dimensional point
of view.  If they wrap the same cycle but with opposite orientations, we
have a brane-antibrane configuration, where supersymmetry is completely
broken at the string scale \cite{bs95}.  But for a generic configuration
with topologically different cycles, there is a non-zero relative angle,
let us call it $\theta$, in the range $0\leq \theta \leq \pi$, and the
squared supersymmetry-breaking mass scale becomes proportional to
$2\theta/(2\pi \alpha')$, for small angles.  The case we are considering
can be understood as an intermediate case between the supersymmetric
parallel branes ($\theta =0$) and the extreme brane-antibrane pair
($\theta =\pi$), with the angle playing the role of the smooth
supersymmetry breaking parameter in units of the string length.

\begin{figure}[htb]
\vspace*{-0.7cm}
\begin{center}
\includegraphics[width=12cm]{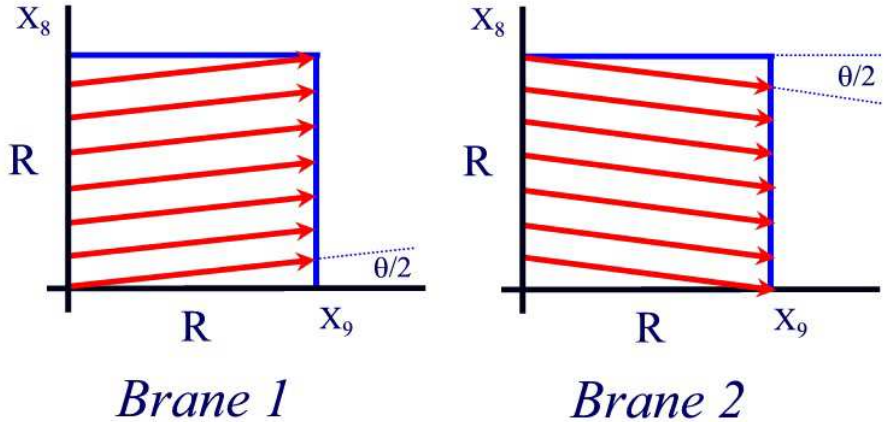}
\vspace*{-1cm}
\caption{The wraping of the two D4-branes around the compactified space
at an angle $\theta/2$ with opposite orientations. The relative angle
between the two branes is $\theta$. If the branes wrap many times along
one direction, the angle $\theta$ can be quite small.}
\end{center}
\label{braneangle}
\end{figure}

Notice that this configuration does not satisfy the R-R tadpole
conditions \cite{pc88}. These conditions state that the sum of the
homology cycles where the D-branes wrap must add up to zero. In our case
these conditions do not play an important role, since we can take a
brane, with the opposite total charge, far away in the transverse
directions. This brane will act as an {\em expectator} during the
dynamical evolution of the other two branes.  Also, since the
configuration is non-supersymmetric, there are uncancelled NS-NS
tadpoles that should be taken into account
\cite{fs86,dm00,bf00,p98}. They act as a potential for the internal
metric of the manifold, \eg the complex structure of the last two torus
in the 8-9 plane, and for the dilaton. Along this paper we will consider
that the evolution of these closed string modes is much slower than the
evolution of the open string modes.

Each brane is located at a given point in the two planes determined by
the compact directions $\{x^4,...,x^7\}$; let us call these points
$y_i$, $i=1,2$. In the supersymmetric ($\theta=0$) configuration there
is no force between the branes and they remain at rest. When $\theta
\not=0$, a non-zero interacting force develops, attracting the branes
towards each other. From the open string point of view, this force is
due to a one-loop exchange of open strings between the two branes.
The coordinate distance between the two branes in the compact space,
$y^2 = |y_1 - y_2|^2$, plays the role of the inflaton field, whose
vacuum energy drives inflation. At tree level, the vacuum energy is
provided by the D4-brane tension $T_4$ and the brane length $L$ around
the 1-cycle, $V_0 = 2\,T_4\,L$.  At one loop, the potential is given
by the zero-point vacuum energy,

\begin{eqnarray}\nonumber
\Vloop &=& \sum_i \int {d^3k\over(2\pi)^3}\,{\hbar\omega_k^i\over2}\\[2mm]
&=& {1\over64\pi^2}\sum_i(-1)^{F_i}\, m_i^4\,\ln m_i^2 \label{CWpot}\\[2mm]
&=& {-1\over(4\pi)^2}\int_0^\infty {dt\over t^3}\,\sum_i (-1)^{F_i} 
e^{-t\,m_i^2}\label{Vloop}
\end{eqnarray}
where the sum goes over all string states of mass $m_i$. This
expression is nothing but the Coleman-Weinberg potential for the low
energy effective field theory. The fact that $\sum_i (-1)^{F_i}\,
m_i^{2n} = 0,\ (n=1,2,3)$, ensures that (\ref{CWpot}) is finite, as it
should be, since supersymmetry is being spontaneously broken.

We can now compute the full potential at 1-loop order, following
Ref.~\cite{p98}
\begin{equation}\label{potD4}
\Vloop(y_i,\theta) = - \int_0^\infty {dt\over t}
(8 \pi^2 \alpha'\,t)^{-2} \sum_{n_i} e^{- {t\over2\pi\alpha'} 
\sum_i (y_i + n_i 2\pi R_i)^2} Z(\theta,t)\,,
\end{equation}
which splits into three terms: The first term contains the integration
over momenta in the (3+1) non-compact dimensions. The second term is
the sum over winding modes in the compact transverse dimensions. For
small distances $y$ compared with the size of the compactified space,
it becomes
\begin{equation}\label{winding}
\sum_{n_i} e^{- {t\over2\pi\alpha'} \sum_i (y_i + n_i 2\pi R_i)^2}
\ \stackrel{y_i \ll 2\pi R}{-\!\!\!-\!\!\!-\!\!\!\longrightarrow}\ 
e^{- {t\,y^2\over2\pi\alpha'}}\,,
\end{equation}
with $y^2=\sum_i y_i^2$ the coordinate distance between the branes in
the compact space.  Branes know about this space only through their
winding modes.  If the initial condition satisfies $y_i \ll 2\pi R$,
the winding modes are so massive that they decouple. Once the branes
start to fall towards eachother we can ignore the presence of the
compact space thereafter. Of course, we will be assuming that the
actual size ($R$) of the compact space becomes fixed by some yet
unknown mechanism, and that the branes are thus free to move within
this space.

The last term is the string partition function,
\begin{equation}\label{Zeta}
Z(\theta,t) = {\Theta^4_{11}(i \theta t/2\pi,it)\over
i\,\Theta_{11}(i\theta t/\pi,it)\,\eta^9(it)} \,,
\end{equation}
which contains a sum over all open string modes propagating between
the branes. The $\eta(\tau)$ and $\Theta_{11}$ are the Dedekind and
the Jacobi elliptic functions that come from the bosonic and fermionic
string modes, respectively.

At large distances compared with the string scale, $l_s \ll y \ll 2\pi
R$, the terms dominating the integral in Eq.~(\ref{potD4}) come from
the limit $t\to 0$,
\begin{equation}\label{Zeta0}
Z(\theta,t) \ \stackrel{t\to 0}{\longrightarrow}\
4\,t^3\,\sin^2{\theta\over2}\,\tan{\theta\over2}\,.
\end{equation}

The interbrane potential thus becomes
\begin{equation}\label{Vytheta}
V(y, \theta) = V_0 + \Vloop =
2\,T_4\,L - {\sin^2{\theta\over2}\,\tan{\theta\over2}\over8\pi^3
\alpha'\,y^2}\,,
\end{equation}
with $\theta\ll1$ fixed. Therefore, the 4-dimensional action is
given by
\begin{equation}
S = \int d^4x\,\sqrt{-g}\left[\half M_P^2\,R - \half T_4\,L\,
(\partial y)^2 - V(y)\right]\,,
\end{equation}
with a Planck mass \ $M_P^2 = M_s^2(2\pi R\,M_s)^6\,g_s^{-2}$.
The canonically normalized field we will call the inflaton is 
\ $\phi = (T_4\,L)^{1/2}\,y$, which has an effective potential at
large distances given by
\begin{eqnarray}\label{Vphi}
V(\phi) &=&  M^4\left(1 - \beta\,{M_s^2\over\phi^2}\right)\,,\\
M^4 &\equiv&  2T_4\,L = {M_s^4(M_s\,L)\,g_s^{-1}\over32\pi^4}\,,\\
\beta &\equiv& {\sin^2{\theta\over2}\,\tan{\theta\over2}\over16\pi^3}
\simeq {\theta^3\over128\pi^3} \ll 1\,.\label{beta}
\end{eqnarray}
This potential has the expected form, with the right power of the
distance, as would arise from Gauss law.

At short distances, $y\leq l_s$, there is a string mode that acquires
a negative mass squared, signalling the end of inflation through the
tachyon condensation. If $\theta\neq0$, there exists a non-trivial cycle
that minimizes the volume, and hence the energy, of the system and the
two branes become unstable, see Fig.~6.

\begin{figure}[htb]
\begin{center}
\includegraphics[width=12cm]{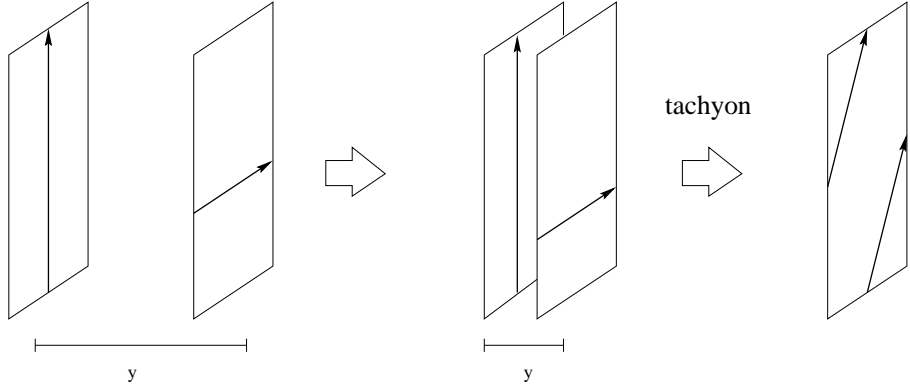}
\caption{Two D4-branes attract each other to decay in the last step to a
bound state. Inflation will take place when the two branes are far away
if the angle between the two branes is small enough. We have chosen here
$\theta=\pi/2$ for pictorical purposes. In fact the branes are at an
angle $\theta\ll1$, so we would expect the final brane to be wrapped
around the corresponding cycle.}
\end{center}
\label{choque}
\end{figure}

From the low energy effective theory, it signals the appearance of
the string tachyon, with mass
\begin{equation}
m_T^2 = {y^2\over4\pi^2{\alpha'}^2} - {\theta\over2\pi\alpha'}\,,
\end{equation}
which defines the bifurcation point, $y = y_c = \sqrt{2\pi\theta}\
l_s$, below which the mass squared is negative $m_T^2<0$, and the
tachyon triggers the end of inflation, in a way completely analogous
to that which occurs in hybrid inflation~\cite{hybrid}, see
Fig.~7.

\begin{figure}[htb]
\begin{center}
\includegraphics[width=8cm]{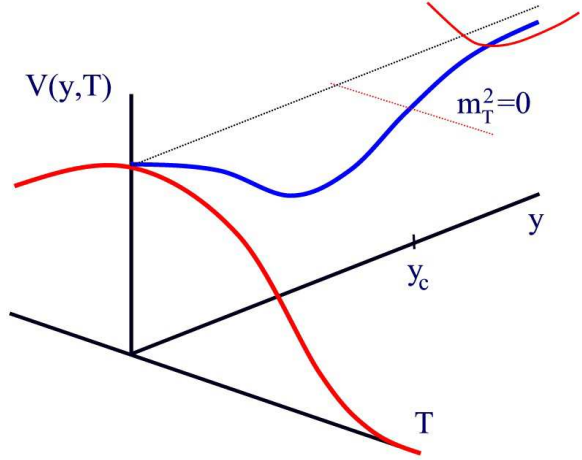}
\caption{A sketch of the inflaton-tachyon potential
$V(T,y)$. The flat region corresponds to the inflationary regime.
The red lines correspond to the tachyon potential at large 
distances $y\gg y_c$, and at $y=0$. Note that it approaches
asymptotically the tachyon vev at infinity.}
\end{center}
\label{inflatontachyon}
\end{figure}

\section{Brane inflation}

Let us now use the inflaton potential (\ref{Vphi}) to derive
phenomenological bounds on the parameters of the model. Note that a
small angle $\theta \ll \pi$ ensures $\beta\ll 1$ and therefore a flat
potential.  Also, $\theta \ll \pi$ implies $y_c \ll l_s$, so that we
may have enough number of \efolds before the string tachyon triggers
the end of inflation.

The number of \efolds and the slow-roll parameters can be computed
as~\cite{Lindebook,LythRiotto,JGB}
\begin{eqnarray}
\label{Ne}
N &=& {1\over M_P^2}\,\int {V\,d\phi\over V'(\phi)} = {1\over8\beta}
\left({M_s^2\over M_P^2}\right)^2\,{\phi^4\over M_s^4} \gg 1 
\hspace{1cm} {\rm for} \ \beta\ll 1\,,\\[2mm] 
\label{epsilon}
\epsilon &=& {M_P^2\over2}\left({V'(\phi)\over V(\phi)}\right)^2 
\simeq {1\over 32\,N^2}\,{\phi^2\over M_s^2} \ll 1 \,,\\[2mm] 
\label{eta}
\eta &=& M_P^2\,{V''(\phi)\over V(\phi)} = -\,{3\over4N}\,,\\[2mm]
\xi &=&  M_P^4\,{V'\,V'''(\phi)\over V^2(\phi)} = {3\over4N^2}\,.
\label{xi}
\end{eqnarray}
Substituting $N\simeq58$, we find the scalar tilt and its running as
\begin{eqnarray}\label{tilt}
n_s &=& 1 + 2\eta - 6\epsilon \ \simeq \ 1 -\,{3\over2N} \simeq 0.974\,,\\[2mm]
{d\,n_s\over d \ln k} &=& -2\xi -{2\over3}\Big[(n_s-1)^2-4\,\eta^2\Big]
\ \simeq -\,{3\over2N^2} \simeq -4\times10^{-4}\,,
\end{eqnarray}
which constitute specific predictions, that should be compared with
observations of the anisotropies of the microwave background.  For the
moment, COBE~\cite{COBE} and Boomerang~\cite{Boomerang}, as well as
the recent results of WMAP~\cite{WMAP} seem to be in agreement with
these predictions, at the 2$\sigma$ level.  Furtheremore, the observed
amplitude of temperature fluctuations fixes the size of the
intersecting angle between branes,
\begin{eqnarray}
\delta_H &=& {1\over5\pi\sqrt3} {V^{3/2}\over V'\,M_P^3} =
{N^{3/4}2^{-1/4}\over5\sqrt3\pi^3}{g_s\,(M_s\,L)^{1/2}\over
\beta^{1/4}(2\pi\,R\,M_s)^{9/2}} = 1.9\times10^{-5}\,,\\[2mm]
\beta^{1/4} &=& 3.3\times10^3\,{g_s\,(M_s\,L)^{1/2}\over
(2\pi\,R\,M_s)^{9/2}}\,.
\end{eqnarray}
These constrain fix a few scales. For instance, the value of the
inflaton field becomes
\begin{equation}
{\phi_*\over M_s} = (2\beta N)^{1/4}\,(M_s\,2\pi R)^{3/2}\,
g_s^{-1/2} = 1.5\times10^4\,
{(M_sLg_s)^{1/2}\over(M_s\,2\pi R)^3}\,,
\end{equation}
where the asterisc denotes the time when the present horizon-scale
perturbation crossed the Hubble scale during inflation, 58 $e$-folds
before the end on inflation. In terms of the distance between branes,
\begin{eqnarray}
{y_*\over l_s} &=& 4\pi^2\left({g_s\over M_s\,L}\right)^{1/2}
{\phi_*\over M_s} = 6\times10^5\,{g_s\over(M_s\,2\pi R)^3}\,, \\
{y_*\over 2\pi R} &=& {1\over M_s\,2\pi R}\,{y_*\over l_s} =
6\times10^5\,{g_s\over(M_s\,2\pi R)^4}\,, 
\end{eqnarray}
independent of $M_s\,L$. For a possible choice of compactification
radius of order $2\pi R\,M_s = 60$, string coupling within
perturbation theory, $g_s = 0.1$, and a reasonable wrapping of the
brane around the $T^2$ torus, $M_s\,L\simeq200$ or $L\simeq 40R$, we
obtain $\theta=1.6\times10^{-3}$, which determines $y_* = 3\,l_s =
(2\pi R)/14$ for the initial stage of inflation, while $y_c =
0.1\,l_s$ at the bifurcation point. Note that the model is thus very
robust with respect to initial conditions: the condition $l_s \ll y
\ll 2\pi R$ for a wide range of models ensures that inflation can
take place with enough number of \efolds to solve the horizon and
flatness problems.

For this particular value of $2\pi R\,M_s = 60$, we obtain a string
scale, a Hubble rate and a radius of compactification
\begin{eqnarray}
M_s &=& {M_P\,\,g_s\over(2\pi R\,M_s)^3} \simeq 
9\times10^{19} \ {\rm GeV}\,, \\
M &=& (2M_sL\,g_s^{-1})^{1/4}{M_s\over2\pi} \simeq 
1\times10^{13} \ {\rm GeV}\,, \\
H &=& {M^2\over\sqrt3M_P} \simeq 3\times10^7 \ {\rm GeV}\,, \\
R^{-1} &=& 2\times10^{12} \ {\rm GeV}\,.
\end{eqnarray}
We may ask whether our approximation of using a 4-dimensional
effective theory is correct. For that we need to check that the
size of the compactified space is much smaller than the Hubble
radius $H^{-1}$ of the 4D theory during inflation,
\begin{equation}
R/H^{-1} = {\sqrt2\over(2\pi)^3}\,
{(M_sLg_s^{-1})^{1/2}\over(M_s\,2\pi R)^2} = 1.5\times10^{-5} \ll 1\,,
\end{equation}
so we are safely within an effective 4D theory.

Finally, let us study the constraints coming from the stochastic
background of gravitational waves produced during inflation, with
amplitude and tilt,
\begin{eqnarray}\label{Pg}
{\cal P}_g^{1/2} &=& {\sqrt2\over\pi}{H\over M_P}   < 10^{-5} 
\hspace{2cm}  {\rm  CMB\ bound}\,, \\
n_t &=& {d\ln{\cal P}_g(k)\over d\ln k} \simeq - 2\epsilon  
\ll 1  \hspace{1cm}  {\rm  (not\ yet\ observed).}
\end{eqnarray}
Substituting the corresponding expressions, we obtain
\begin{equation}
{\cal P}_g^{1/2} = {4\over\sqrt3(2\pi)^3}\,
{(M_s L\,g_s^{-1})^{1/2}\over(M_s\,2\pi R)^6} = 2\times10^{-12}\,,
\end{equation}
which is indeed well below the present bound (\ref{Pg}), and we
can  safely ignore the gravitational wave background.

\section{Geometrical interpretation of brane inflation parameters}

Here we will briefly give a geometrical interpretation of the number
of \efolds and the slow-roll parameters in our model. The epsilon
parameter (\ref{epsilon}) is in fact the relative velocity ($v=\dot
y$) of the branes in the compact dimensions. Since $\dot\phi^2=
M^4v^2/2$ and $3H^2=M^4/\Mp^2$, we have
\begin{equation}
\epsilon=-{\dot H\over H^2}={\dot\phi^2\over2\Mp^2H^2}={3\over4}\,v^2\,.
\end{equation}

The number of \efolds (\ref{Ne}) can then be interpreted as the
geometric distance between the branes in the compactified space,
\begin{equation}
N = \int{d\phi\over \Mp\sqrt{2\epsilon}}=\int H\,{dy\over v}\,.
\end{equation}
Finally, the eta parameter (\ref{eta}) is the acceleration of the branes
with respect to each other due to an attractive potential of the type
$V(y)\propto y^{2-d_\bot}$, coming from Gauss law in $d_\bot$ transverse
dimensions,
\begin{equation}
\eta = - {d_\bot-1\over d_\bot\,N}\,,
\end{equation}
which {\em only} depends on the dimensionality of the compact space
$d_\bot$. Note that the spectral tilt of the scalar perturbations
(\ref{tilt}) therefore depends on both the velocity and acceleration
within the compact space, and is very small for only a spontaneous
supersymmetry breaking, which makes the branes approach eachother very
slowly, driving inflation and giving rise to a scale invariant
spectrum of fluctuations.

\begin{figure}[htb]
\begin{center}
\includegraphics[width=8cm]{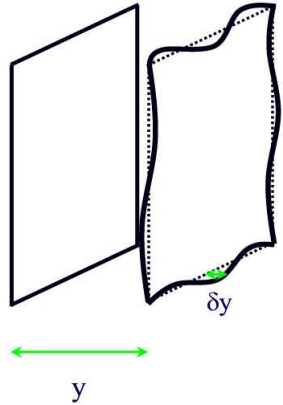}
\caption{The inflaton field $y$ is interpreted as the distance between
the two branes. Quantum fluctuations of this field will give rise upon
collision to density perturbations on comoving hypersurfaces.  These
fluctuations will be later observed as temperature aniso\-tro\-pies in
the microwave background.}
\end{center}
\label{fluctuations}
\end{figure}

In fact, the induced metric perturbations in our (3+1)-dimensional
universe can be understood as arising from the fact that, due to 
fluctuations in the approaching D4-branes, inflation does not end at the
same time in all points of our 3-dimensional space, see
Fig.~8, and the gauge invariant curvature perturbation
on comoving hypersurfaces, ${\cal R}_k = \delta N_k = H\,\delta y_k/v$,
is non-vanishing, being later responsible for the observed spectrum
of temperature anisotropies in the microwave
background~\cite{COBE,Boomerang,WMAP}.

\section{Reheating the Universe after inflation}

Reheating is the most difficult part in inflationary model building
since we don't know to what the inflaton couples to. Eventually one
hopes the vacuum energy during inflation will go into relativistic
degrees of freedom and everything will finally thermalise so that the
hot Big Bang follows. One thing we know for sure is that the
universe must have reheated before primordial nucleosynthesis ($T_{\rm
rh} > 1$ MeV), otherwise the light element abundances would be in
conflict with observations. However, since the scale of inflation is
not yet determined observationally, we are allowed to consider
reheating the universe just above an MeV.

We will briefly describe reheating after brane inflation. The details
have not been worked out yet, but we can give here a succint account
of what should be expected. Brane inflation ends like hybrid
inflation, still in the slow-roll regime, when the string-tachyon
becomes massless and the tachyon condenses at the true minimum. In
Ref.~\cite{SSB} it was shown that this typically occurs very fast,
within a time scale of order the inverse curvature of the tachyon
potential, $t_* \sim m_T^{-1} \equiv (\theta/2\pi\alpha')^{-1/2}$.
From the point of view of the low energy effective field theory
description, this is seen as the decay rate (per unit time and unit
volume) or the imaginary part of the one-loop energy density
(\ref{V0infty}).

\begin{figure}[htb]
\begin{center}
\includegraphics[width=12cm]{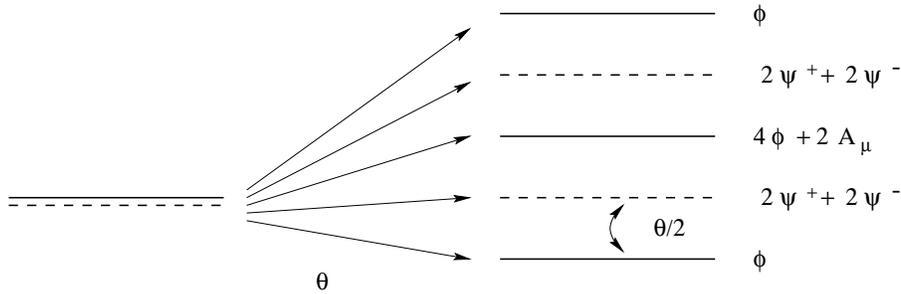}
\caption{Splitting of the mass
spectrum of the ${\cal N} = 4$ Yang-Mills supermultiplet when the angles
are non-vanishing or, in the T-dual picture, degenerancy splitting for
the first Landau level.  $\psi^{\pm}$ are the chiral fermions and $\phi$
are the scalars.}
\end{center}
\label{split}
\end{figure}

Let us calculate the low energy degrees of freedom in the brane after 
the collision. This can be extracted from the partition function
(\ref{Zeta}),written here in terms of infinite products,
\begin{eqnarray}\label{Ztheta}
Z(\theta,t) &=& {(z-1)^4\over z}\,\Big(\sum_{n=0}^\infty z^{2n}\Big)\,
\prod_{m=1}^\infty {(1-q^m z)^4(1-q^m z^{-1})^4\over(1-q^m)^6
(1-q^m z^2)(1-q^m z^{-2})} \,,\\
q &=& e^{-2\pi t}\,, \hspace{1cm} z = e^{-\theta t}\,.
\end{eqnarray}
The first factor, $(1-z)^4/z = z^{-1}-4+6z-4z^2+z^3$, gives precisely
the lowest lying ${\cal N}=4$ supermultiplet, including the tachyon,
see Table~1 and Fig.~9, with the correct
multiplicity and (bosonic/fermionic) sign. The fact that $\sum_i
(-1)^{F_i}\,m_i^{2n} = 0,\ (n=1,2,3)$, ensures that (\ref{CWpot}) is
finite, as it should be, since supersymmetry is only broken
spontaneously by the angle $\theta$.

\begin{table}[h]
\label{mass1}
\hspace{3cm}
\begin{tabular}{|c|c|} \hline
Field & $2\pi\alpha'm^2$ \\
\hline
\hline
1 scalar &  $y^2/(2\pi\alpha')+3 \theta$ \\
\hline 
2 massive fermions &   $y^2/(2\pi\alpha')+2\theta$ \\
\hline 
3 scalars &   $y^2/(2\pi\alpha')+\theta$ \ \\
\hline 
1 massive gauge field &   $y^2/(2\pi\alpha')+\theta$ \ \\
\hline 
2 massive fermions (massless for $y=0$) &   $y^2/(2\pi\alpha')$ \ \ \\
\hline 
1 scalar (tachyonic for $y=0$) &   $y^2/(2\pi\alpha') - \theta$ \ \\
\hline 
\end{tabular}
\caption{The mass spectrum of the ${\cal N} = 4$ supermultiplet.
Notice that the first scalar will be tachyonic if the distance between
the two branes is smaller than $y_c^2 = 2\pi\alpha'\theta$, \ie the
two-brane system becomes unstable.}
\end{table}

The next factor in (\ref{Ztheta}) is the sum $\sum_{n=0}^\infty z^{2n}
= 1 + z^2 + z^4 + z^6 + \dots$, which corresponds to the Landau levels
induced, in the dual picture, by the supersymmetry breaking flux
associated to the angle $\theta$. They give, at any order $N$, a
supermultiplet with $\sum_i (-1)^F\,m_i^{2n} = 0,\ (n=1,2,3)$, so they
still provide a finite one-loop potential (\ref{CWpot}).  For a given
supersymmetry-breaking angle $\theta$, one should include in the low
energy effective theory the whole tower of Landau levels up to
$N=1/\theta$.

Finally, one could include also the first low-lying string states,
whose masses are determined by the expansion of the infinite products
in (\ref{Ztheta}),
\begin{equation}
Z(\theta,t) = {(z-1)^4\over z}\,\Big(\sum_{n=0}^\infty z^{2n}\Big)\,
\Big[1 + {(1-z)^4\over z^2}\,q^2 + {(1-z)^4(1+7z^2+z^4)\over z^4}\,q^4
+ \dots \Big]\,.
\end{equation}
Their structure still comes in ${\cal N}=4$ supermultiplets, so they
again give a finite Coleman-Weinberg potential.

We will use the whole tower of Landau levels in the effective field
theory to connect the one-loop potential at short distances, with the
full string theory one-loop potential coming from exchanges of the
massless string modes at large distances, responsible for
inflation. This connection will be essential for the latter stage of
preheating and reheating, because it will provide the low energy
effective masses, and the couplings between the inflaton field $y$ and
the effective fields living on the D4-brane.

The potential (\ref{Vloop}) is finite if there are no massless or
tachyonic fields.  When the tachyon appears, \ie at distances smaller
tham $y_c$, there is an exponentially divergent amplitude for $m_i=0$.
A possible strategy to attach physical meaning to this divergence is
to analytically continue the potential (\ref{potD4}) in the complex
$y$-plane.  After the continuation, there is a logarithmic branch
point at $y=y_c$. In this way, we get rid of the divergence and the
potential develops a non-vanishing imaginary part for $y<y_c$, which
signals the instability of the vacuum.

\begin{figure}[htb]
\vspace*{4.5cm}
\begin{center}
\hspace*{-5cm}
\includegraphics[width=6cm]{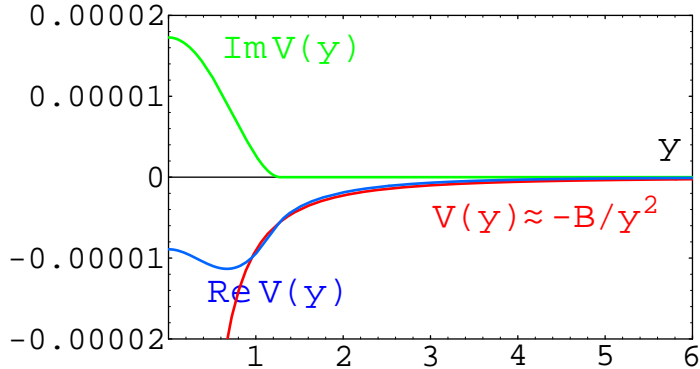}
\vspace*{-2cm}
\caption{The attractive inflaton potential between two D4-branes. The
red line corresponds to the string-derived potential (\ref{potD4}); in
blue is the Real part of the Coleman-Weinberg field-theory limit of the
same potential (\ref{Vloop}), and in green is the Imaginary part, which
is non-zero only when the tachyon condenses, at $y^2 <
2\pi\alpha'\theta$. We have chosen here $\theta=\pi/12$. The potential
$V(y)$ and the distance $y$ are both in units of $\alpha'=1$.}
\end{center}
\label{potential}
\end{figure}

We have plotted in Fig.~10 the attractive potential $V(y)$
between two D4-branes at an angle $\theta=\pi/12$. The large distance
behaviour $y^2 \gg \alpha'$ is determined from the supergravity
amplitude (\ref{potD4}), in red, while the short distance potential is
obtained from the Coleman-Weinberg potential corresponding to the
lowest-lying effective fields, which fall into ${\cal N} = 4$
supermultiplets. The real part of the Coleman-Weinberg potential is
drawn in blue in Fig.~10, while the imaginary part is in
green.

In the field theory limit ($q\to0$ and $z\neq0$) we can consider the
infinite tower of Landau levels, while ignoring the string levels, \ie
$Z(\theta,t) = (z-1)^4\,\sum_{n=0}^\infty z^{2n-1}$.  The finite low
energy effective potential at zero separation $y=0$ becomes~\cite{brz01}
\begin{equation}\label{V0infty}
V(0,\theta) = {i\pi-1.70638\over32\pi^2}\,
\left({\theta\over2\pi\alpha'}\right)^2\,.
\end{equation}
The imaginary part can be related, through the optical theorem, with
the rate of decay of the false vacuum towards the minimum of the
tachyon potential,
\begin{equation}
\Gamma_\phi = (2\pi R)^3\,{\rm Im}\,\Vloop \simeq 3\times10^{-5}\,M_s
\simeq 2.7\times10^8\ {\rm GeV}\,.
\end{equation}
This implies a (perturbative) reheating temperature
\begin{equation}
\trh \simeq 0.1\,\sqrt{\Gamma_\phi\,M_p} 
\simeq 3\times10^{12}\ {\rm GeV}\,,
\end{equation}
which coincides with the reheating temperature that would have
resulted if all the false vacuum energy would have been converted into
radiation soon after symmetry breaking. Note that since the rate of
expansion during inflation $H=M^2/\sqrt3\Mp$ is much smaller than the
mass scale $m_T$, the false vacuum energy density $E_0=2T_pL=M^4$ is
not diluted by the expansion before reheating, as typically occurs in
chaotic inflation, and all of it is converted into radiation,
reheating the universe to a temperature
\begin{equation}
T_{\rm rh} = \Big({30\over\pi^2g_*}\Big)^{1/4}\,M \simeq
1\times10^{12} \ {\rm GeV}\,,
\end{equation}
where we have assumed $g_* \sim 10^3$ for the number of relativistic
degrees of freedom at reheating.

However, the actual process of reheating is probably very complicated
and there is always the possibility that some fields may have their
occupation numbers increased exponentially due to parametric
resonance~\cite{KLS} or tachyonic preheating~\cite{SSB}. Moreover, a
significant fraction of the initial potential energy may be released
in the form of gravitational waves, which will go both to the bulk and
into the brane. One must be sure that the bulk gravitons do not reheat
at too high a temperature, because their energy does not redshift
inside the large compact dimensions (contrary to our (3+1)-dimensional
world, where radiation redshifts with the scale factor like $a^{-4}$),
and they could interact again with our (presently cold) brane world
and inject energy in the form of gamma rays, in conflict with present
bounds from observations of the diffuse gamma ray
background~\cite{ADD}.  Fortunately, since the fundamental
gravitational scale, $\Mp$, in this model is large enough compared
with all the other scales, the coupling of those bulk graviton modes
to the brane is suppressed, and thus we do not expect in
principle any danger with the diffuse gamma ray background~\cite{ADD},
but a detailed study remains to be done.

\section{Conclusions}

We have shown how string theory provides an interesting realization of
hybrid inflation, in the context of D-branes wrapped around the
compactification scale, where the inflaton is the interbrane
separation. The small parameter needed for sufficient inflation and a
small amplitude of fluctuations in the CMB is due here to the
spontaneous breaking of supersymmetry via a small angle between the
branes, which induces a very flat potential for the inflaton.

This model gives a geometrical realization of inflation which is very
robust with respect to initial conditions in the compact space.
Moreover, the slow-roll conditions can be interpreted as (geometrical)
conditions on the relative velocity and acceleration of the branes
towards eachother. The number of \efolds is directly related to the
distance between branes, and metric perturbations, that later give
rise to temperature anisotropies, are nothing but fluctuations on the
interbrane distance.

The model makes very specific predictions:
For a concrete compactification radius in units of the string
scale, $2\pi R=60\,l_s$, we find a mass scale $M \sim M_s \sim 10^{13}$
GeV. Moreover, the radius of compactification turns out to be $R^{-1}
\sim 10^{12}$ GeV. The scalar tilt is $n_s = 0.974$, with negligible
scale dependence, $d n_s/d\ln k = -3\times10^{-4}$, and there is an
insignificant amount of tensor modes or gravitational waves.

The reheating of the universe occurs through the conversion of all
the false vacuum energy during inflation into radiation, as the
tachyon condenses and is driven towards its vev. A detailed account
of this process is still lacking.

\vskip5mm

\section*{ACKNOWLEDGEMENTS}
I would like to thank the organizers of the Davis Meeting on Cosmic
Inflation for a very warm and friendly atmosphere. It's also a
pleasure to thank my friends and collaborators Ra\'ul Rabad\'an and
Frederic Zamora, for sharing with me their insight about branes in
string theory. This work was supported by a CICYT project FPA2000-980.


\end{document}